\documentstyle[12pt,epsfig]{article}

\begin{document}
\begin {center}
{\bf {\Large
Modification of $\Delta (1232)$ in the neutrino nucleus reaction } }
\end {center}
\begin {center}
Swapan Das  \\
{\it Nuclear Physics Division,
Bhabha Atomic Research Centre  \\
Mumbai-400085, India }

{\it Homi Bhabha National Institute,  \\
Anushakti Nagar, Mumbai-400094, India }
\end {center}

\begin {abstract}
The remarkable observations in the pion induced and charge exchange
reactions on a nucleus are the significant shift and broadening of the
$\Delta (1232)$-peak relative to free pion nucleon scattering.
For
the forward going massless leptons, the weak process $(\nu_l,l)$ on a
nucleon, according to Adler's partially conserved axial current theorem,
is connected to the pion nucleon scattering.
This
mechanism is also applicable to the nucleus. Therefore, the modification
of the $\Delta$-peak in a nucleus can be seen in the $(\nu_l,l)$ reaction
since it is connected to the pion nucleus scattering.
To
investigate this issue quantitatively, the double differential cross
sections of the forward going ejectile $l$ energy distribution in the
$(\nu_l,l)$ reaction are calculated in the $\Delta$-excitation region
for both proton and nucleus.
The
measured pion nucleon and pion nucleus scattering cross sections in the
quoted energy region are used in these reactions as input.
Since
the $(\nu_l,l)$ reaction is connected to the pion induced scattering, the
features of the previous reaction is shown analogous to those of the later.
\end {abstract}

Keywords: $\Delta (1232)$ production, Adler theorem,
neutrino nucleus reaction

PACS number(s): 25.30.-c; 25.30.Pt 

\medskip
\medskip

\section{Introduction}

In the inclusive pion nucleus scattering, the sizeable broadening and shift
of the $\Delta (1232)$-peak have been found in the pion energy distribution
spectrum relative to the pion scattering on the free nucleon
\cite{bion, clou}.
The
origin of the quoted modification of $\Delta$-isobar has been understood
by studying the exclusive and semi-exclusive channels of the pion nucleus
scattering, such as, elastic and inelastic pion production, pion absorption,
... etc, \cite{ashe}.
The
coherent multiple scattering of pion, mediated by the delta-hole propagation
\cite{oset, hira}, in the nucleus leads to the large shift and broadening
of the peak of the cross section in the elastic pion nucleus scattering.
In
fact, the inelastic and reaction channels of the pion nucleus scattering
are also equally significant throughout the $\Delta$-excitation region
\cite{ashe}.
The
$\Delta$-resonance produced in the pion nucleus scattering strongly couples
to the quasi-free $(\pi,\pi N)$ and two-body absorption $(\pi,NN)$ channels
which also modify the hadronic parameters of the $\Delta$-isobar in the
nucleus.
These channels, except $\pi N$, do not occur for proton.

The modification of the $\Delta (1232)$-peak has been also seen in the
measured ejectile energy distribution spectrum in the inclusive
$(^3\mbox{He},t)$ reaction on the nucleus relative to proton \cite{cont}.
This
phenomenon is also found in other inclusive reactions, e.g., $(p,n)$
\cite{bjork}. However, the position of the $\Delta$-peak in
this reaction (unlike that in the pion nucleus scattering) does not show
much dependence on the target mass number.
To
understand the origin of $\Delta$-peak shift in the inclusive reactions,
various models (describing the $\Delta$-dynamics in the nucleus)
were proposed \cite{esben}.

To resolve the above issue, the exclusive/semi-exclusive experiments on the
charge exchange reactions (i.e., $(p,n)$ \cite{chiba} and $(^3\mbox{He}, t)$
\cite{henn}) were done in the $\Delta (1232)$-resonance region.
In
these experiments, the ejectile energy/momentum was measured with the
charged particles (i.e., $\pi^+$, $\pi^+p$, $pp$, ... etc) detected in
coincidence.
According
to these measurements, the shift of $\Delta$-peak in the nucleus (seen in
the inclusive measurements) arises because of the $pp$ emission and coherent
pion production which do not exist for proton. The later is the key
ingredient for the $\Delta$-peak shift in the nucleus.
The
results of the (semi)exclusive measurements illustrate one to one
correspondence with those of the pion nucleus scattering \cite{das}.

In contrast to those found in the pion induced and charge exchange reactions,
the shift of the $\Delta (1232)$-peak in the nucleus is not seen in the
inclusive $\gamma$ induced \cite{aren} and $(e,e^\prime)$ reactions
\cite{conn}.
In
these reactions, the $\Delta$-excitation occurs due to the transverse
$\gamma N\Delta$ coupling but its decay, i.e., $\Delta \to \pi^+p$, proceeds
because of the longitudinal $\pi N\Delta$ coupling. Since these couplings
are orthogonal to each other, the coherent pion production is inhibited in
the photo- and electro- induced nuclear reactions \cite{oset2}.
The
broadening of the $\Delta$-peak, seen in these reactions, arises because of
Fermi motion of nucleon and opening up additional channels, like
$N\Delta \to NN$ \cite{oset, koch}.
In
fact, universality exists for the integrated cross section of the inclusive
$(e,e^\prime)$ reaction because it is same for all nuclei \cite{conn}.

The modification of $\Delta (1232)$ in the nucleus has been taken into
account for a large variety of neutrino nucleus reactions such as
inclusive scattering, incoherent and coherent pion production \cite{singh}.
It
must be mentioned that the cross section of the $(\nu_l,l)$ reaction under
appropriate constraint can be connected to that of the pion induced
scattering using Adler's partially conserved axial current (PCAC) theorem
\cite{adler}.
According
to this theorem, the $(\nu_l,l)$ reaction can be interpreted in
terms of the pion induced reaction at the target, provided the massless
lepton $l$ moving parallel to the incoming massless $\nu_l$.
The
effect of the finite mass of lepton $l$ in the $(\nu_l,l)$ reaction, as
discussed by Adler \cite{adler}, can be taken into account by incorporating
the lepton mass correction factor $C_{lm}$ in the cross section:
\begin{equation}
C_{lm} =
\left [ 1 - \frac{ m^2_l q_0 }{ 2(m^2_\pi E_l + m^2_l q_0) } \right ]^2;
\label{mcrr}
\end{equation}
with $q_0=E_{\nu_l}-E_l$. $m_l$ and $m_\pi$ are masses of the lepton $l$
and pion respectively.
As
illustrated above, the $(\nu_l,l)$ reaction on a nucleus can be connected
to the pion nucleus scattering. Therefore, the distinct $\Delta$
modification in nucleus in the $(\nu_l,l)$ reaction can occur because that
is seen in the pion nucleus scattering.

To disentangle the above issue, the double differential cross section
$ \frac{d\sigma} {dE_{e^-} d\Omega_{e^-}} $ of the forward going ejectile
energy $E_{e^-}$  distribution in the $(\nu_{e^-},e^-)$ reaction, using
Adler's PCAC theorem \cite{adler}, is expressed in terms of the cross
section of the pion induced reaction.
To
calculate $ \frac{d\sigma} {dE_{e^-} d\Omega_{e^-}} $ of the previous
reaction in the $\Delta$-excitation region, the energy dependent measured
cross section of the later reaction in this energy region has been
used as input.
Two
aspects of the considered reaction, stated above, should be emphasized.
(i)
The
$(\nu_{e^-},e^-)$ reaction is preferred since the masses of the leptons in
the $\Delta$-excitation region are negligibly small. It could be more
appropriate (compare to other flavors of neutrino) for the use of Adler
theorem.
The forward going electron $e^-$ is considered. In such case, the outgoing
lepton $e^-$ moves parallel to the incoming neutrino $\nu_{e^-}$. This is
the requirement of Adler theorem.
(ii)
The measured cross section of the pion induced reaction is used as input.
The advantage of it is that all kinds of mechanism for this reaction are
taken into account.

\section{Formalism}

The matrix element describing the weak process $\nu_{e^-}+A \to e^-+X$,
using the phenomenological $(V-A)$ theory for weak interaction \cite{will},
can be written as
\begin{equation}
{\cal M}_{fi} =
\frac{G_W}{\sqrt{2}} l_\mu <X|(V^\mu-A^\mu)|A>.
\label{gwr1}             
\end{equation}
$G_W \simeq 1.15 \times 10^{-5} ~\mbox {GeV}^{-2}$ is the empirical value
of Fermi coupling constant.
$l_\mu$
denotes the leptonic weak current, i.e.,
$ {\bar u}_e \gamma_\mu (1-\gamma_5) u_{\nu_e} $, which can be written as
$l_\mu = {\cal M}(e\nu_e) q_\mu$ for the forward going massless leptons.
$<|{\cal M}(e\nu_e)|^2>$ (which appears in the cross section) is expressed
afterwards.
Henceforth,
the superscript on the electron is dropped in some places. $V^\mu$ and
$A^\mu$ are the hadronic vector and axial vector currents respectively.
$|A>$
is the initial state of the nucleon (or nucleus) where as $|X>$ represents
the final state (except lepton $l$) of the reaction.
Using
Adler theorem for massless leptons, ${\cal M}_{fi}$ in Eq.~(\ref{gwr1})
can be factorized as
\begin{equation}
{\cal M}_{fi} =
-i \frac{G_W}{\sqrt{2}} {\cal M} (e\nu_e) f_\pi {\cal M} (\pi^a A \to X).
\label{gwr2}             
\end{equation}
$f_\pi$ (=0.13 GeV) is the pion weak decay constant.
$ {\cal M} (\pi^a A \to X) $ denotes the matrix element for the
$\pi^a A \to X$ reaction.
In
fact, conserved vector current (CVC) and partially conserved axial current
(PCAC) hypotheses are applied to get this equation.

The double differential cross section $\frac{d\sigma}{dE_ed\Omega_e}$ of
the forward going lepton energy $E_e$ distribution in the 
$(\nu_{e^-},e^-)$ reaction on proton or nucleus, using ${\cal M}_{fi}$
in Eq.~(\ref{gwr2}), can be expressed as
\begin{equation}
\frac{d\sigma}{dE_ed\Omega_e} =
\frac{\pi^2}{(2\pi)^5} \frac{k_e k_\pi}{k_{\nu_e}}
G^2_W
<|{\cal M} (e\nu_e)|^2> f^2_\pi \sigma_t(\pi^a A \to X),
\label{dfx1}
\end{equation}
with
$ <|{\cal M} (e\nu_e)|^2> = $
$ \frac{16 k_{\nu_e} k_e} {(E_{\nu_e} - E_e)^2} $.
$ \sigma_t(\pi^a A \to X) $ is to be replaced by
$ 2 \sigma_t(\pi^+ A \to X) $ for $\pi^+$ scattering \cite{kama}.
Since
the electron mass correction factor, i.e., $C_{lm}$ in Eq.~(\ref{mcrr}), is
found very close to unity in the $\Delta$-excitation region, the above
equation can be used to calculate the cross sections of the
$A(\nu_{e^-},e^-)X$ reaction in the quoted energy region.

\section{Result and Discussions}

To calculate the double differential cross section
$\frac{d\sigma}{dE_ed\Omega_e}$ for the inclusive $(\nu_{e^-},e^-)$ reaction
on proton and nucleus, the measured total cross sections $\sigma_t(\pi A)$
of the $\pi^+$ proton and inclusive $\pi^+$ nucleus scattering are used in
Eq.~(\ref{dfx1}).
$\sigma_t(\pi A)$ for proton in the $\Delta$-excitation region is taken
from Ref.~\cite{barn}, and those for the inclusive pion nucleus scattering
are taken from Ref.~\cite{clou}.
The
calculated results at $E_{\nu_e}=1$ GeV for the ejectile energy $E_e$
distribution are presented in Fig.~\ref{fg1} for proton and $^{12}$C,
$^{27}$Al, $^{56}$Fe nuclei. The energy transfer, i.e.,
$q_0 = E_{\nu_e}-E_e$ shown by the upper scale of the horizontal axis.
Because
of the large broadening of $\Delta$-resonance in nuclei heavier than Fe,
the distinct $\Delta$-peak is not seen in the measured cross section
$\sigma_t(\pi A)$ of those nuclei. Therefore, they are not considered in
the present context.
Fig.~\ref{fg1}
distinctly shows the modification of $\Delta$-isobar in nuclei in the
inclusive $(\nu_{e^-},e^-)$ reaction. The shift (towards higher ejectile
energy) and broadening of the $\Delta$-peak in $^{12}$C nucleus relative
to proton is significantly large (i.e., $\sim 50 - 60 $ MeV) and those
are found to increase with the size of the nucleus.

It should be mentioned that the $(\nu_{e^-},e^-)$ reaction on proton in the
considered energy region proceeds as $\pi^+p \to \Delta^{++} \to \pi^+p$.
The quasi-free proton knock-out in the $(\nu_{e^-},e^-)$ reaction on a
nucleus is analogous to $\pi^+p$ reaction.
Apart
from it, there are many other channels (i.e., exclusive and semi-exclusive
reactions) opened up for the nucleus which cannot occur for proton. The
cross section of the inclusive reaction is the conglomeration of that of the
exclusive and semi-exclusive reactions.
Therefore,
the origin of the $\Delta$-peak shift, shown in Fig.~\ref{fg1}, can be
understood by studying the exclusive and semi-exclusive $(\nu_{e^-},e^-)$
reactions on the nucleus.

To disentangle the above issue, $\frac{d\sigma}{dE_ed\Omega_e}$ is
calculated for the exclusive and semi-exclusive $(\nu_{e^-},e^-)$ reactions
on the nucleus using the measured total cross sections of the
exclusive and semi-exclusive pion nucleus scattering in Eq.~(\ref{dfx1}).
The
total cross section $\sigma_t (\pi A)$ of the pion nucleus scattering in
the $\Delta (1232)$-resonance region is composed of that due to the exclusive
and semi-exclusive channels \cite{ashe}:
$ \sigma_t (\pi A) = \sigma_{t,el} (\pi A)
                    +\sigma_{t,in} (\pi A)
                    +\sigma_{t,ab} (\pi A)
                    +\sigma_{t,cx} (\pi A) $.
The partial cross sections (elaborated afterwards) in this equation are
almost equal to each other in the peak region, except that of the charge
exchange reaction, i.e., $\sigma_{t,cx} (\pi A)$, which is much less
compared to others \cite{ashe}. Therefore, the later is omitted in the
following discussions.

The calculated forward going ejectile energy $E_e$ distribution spectra
in the exclusive and semi-exclusive $(\nu_{e^-},e^-)$ reactions on $^{12}$C
nucleus in the $\Delta$-excitation region are shown in Fig.\ref{fg2}.
$\frac{d\sigma}{dE_ed\Omega_e}$
calculated using the measured elastic pion nucleus scattering cross section
$\sigma_{t,el} (\pi ^{12}\mbox{C})$ \cite{ashe} in Eq.~(\ref{dfx1})
describes that for the elastic (coherent) pion production in the
$(\nu_{e^-},e^-)$ reaction on the quoted nucleus. It is presented by the
dot-dashed curve in this figure.
The
inelastic pion nucleus scattering consists of many final states, e.g.,
$\pi A^*$, $\pi p X$ ... etc. Amongst them, the quasi-free knock-out
reaction in the nucleus (as mentioned earlier) has one to one
correspondence with the pion nucleon reaction.
The
cross section for the inelastic pion production in the quoted reaction
(shown by the dot-dot-dashed curve) is calculated using the measured
total inelastic pion nucleus scattering cross section $\sigma_{t,in}
(\pi ^{12}\mbox{C})$ \cite{ashe} in Eq.~(\ref{dfx1}).
The 
short-dashed curve represents $\frac{d\sigma}{dE_ed\Omega_e}$
calculated using the measured total
pion absorption cross section, i.e.,
$\sigma_{t,ab} (\pi ^{12}\mbox{C})$ \cite{ashe}, in Eq.~(\ref{dfx1}). In
fact, this describes the cross section for $^{12}$C$(\nu_{e^-},e^-)$
reaction where no pion is in the final state.
It
can be added that the dominant contribution to this reaction arises due to
two-nucleon emission in the final state, which proceeds because of the
elementary reaction $N\Delta \to NN$ occurring in the nucleus, i.e.,
two-particle two-hole (2p-2h) excitation in the nucleus in the
$\Delta$-excitation region.
Therefore,
the quoted reaction can be described by the weak longitudinal response
function (for the charge changing neutrino scattering) due to 2p-2h
excitations in the nucleus  in the considered energy region (e.g., see the
calculation due to Ruiz Simo et al., \cite{ruiz}).
For
comparison, $\frac{d\sigma}{dE_ed\Omega_e}$ of the inclusive
$^{12}$C$(\nu_{e^-},e^-)$ reaction is shown by the solid line in
Fig.~\ref{fg2}.
It
is distinctly visible in this figure that the shift of $\Delta$-peak in the
inclusive $(\nu_{e^-},e^-)$ reaction on the nucleus relative to proton
(as shown in Fig.~\ref{fg1}) arises because of the exclusive and
semi-exclusive channels, i.e., the pion production (both elastic and
inelastic) and absorption, which are not possible for proton.

The calculated results in Figs.~\ref{fg1} and \ref{fg2} are presented for
the fixed neutrino beam energy. Since the energy of the incoming
neutrino flux is not fixed in experiments, the energy dependent neutrino
flux averaged double differential cross section is measured. It can be
expressed as
\begin{equation}
\frac{d<\sigma>}{dq_0d\Omega_e} = \frac{ \int dE_{\nu_e} \Phi(E_{\nu_e})
\frac{ d\sigma (E_{\nu_e}) }{ dq_0d\Omega_e} }
{\int  dE_{\nu_e} \Phi(E_{\nu_e})},
\label{dffx}
\end{equation}
where $q_0=E_{\nu_e}-E_e$ is the energy transfer to the nucleus.
$\Phi(E_{\nu_e})$ describes the neutrino energy dependent flux distribution.
The forward energy transfer distribution spectra in the inclusive
$(\nu_{e^-},e^-)$ reaction are calculated using $\Phi(E_{\nu_e})$ due to
MiniBooNE collaboration (see Fig.~27 in Ref.~\cite{agui}).
The
calculated spectra for proton and $^{12}$C nucleus in the
$\Delta$-excitation region are presented in Fig.~\ref{fg3}. This figure
distinctly shows the modification of the $\Delta$-peak in the nucleus,
which corroborates those illustrated in Fig.~\ref{fg1} for fixed neutrino
beam energy.

The modification of $\Delta(1232)$ in the nucleus is shown to occur in the
$(\nu_{e^-},e^-)$ reaction. To look for that in the heavier $(\nu,l)$
reaction, the double differential cross section is calculated for the
$(\nu_{\mu^-},\mu^-)$ reaction on nuclei. The results calculated
at fixed beam energy, i.e., $E_{\nu_\mu}=1$ GeV, for the forward going
ejectile energy $E_\mu$ distribution is shown in Fig.~\ref{fg4}.
This
figure shows that the modification of $\Delta$ in the $(\nu_{\mu^-},\mu^-)$
reaction is qualitatively similar to that in the $(\nu_{e^-},e^-)$ reaction
presented in Fig.~\ref{fg1}. The reduction in the cross section of the
previous reaction, compared with that of the later, occurs because of the
finite mass correction factor, given in Eq.~(\ref{mcrr}), for the lepton
in the finite state.

As mentioned earlier, the energy dependent neutrino flux averaged cross
section can only be measured. If the flux distributions of $\nu_e$ and
$\nu_\mu$ neutrinos are different, the shape of the flux averaged cross
sections of $(\nu_{e^-},e^-)$ and $(\nu_{\mu^-},\mu^-)$ reactions may not
be similar.
To
explore it for the MiniBooNE flux distribution (see Fig.~27 in
Ref.~\cite{agui}), the $\nu_{\mu^-}$ flux averaged differential cross
section $\frac{d<\sigma>}{dq_0d\Omega_\mu}$ for the forward going $\mu^-$
in the $(\nu_{\mu^-},\mu^-)$ reaction is calculated following
Eq.~(\ref{dffx}), and the calculated results are presented in Fig.~\ref{fg5}.
The
figures \ref{fg3} and \ref{fg5} show that there is not much difference
between the shape of the MiniBooNE flux averaged cross sections of 
$(\nu_{e^-},e^-)$ and $(\nu_{\mu^-},\mu^-)$ reactions,
though
the energy dependent flux distributions of $\nu_{e^-}$ and $\nu_{\mu^-}$
neutrinos are quite different \cite{agui}.
The
shape of $\frac{d<\sigma>}{dq_0d\Omega_\mu}$ is narrower compared with
that of $\frac{d<\sigma>}{dq_0d\Omega_e}$, and the  previous cross
section is smaller than the later. It arises, as stated earlier, because
of the mass correction factor incorporated for the lepton in the final
state.

\section{Conclusions}

The double differential cross sections of the forward going lepton $l$
energy distribution in the $(\nu_l,l)$ reactions on proton and nuclei
have been calculated to look for the modification of $\Delta (1232)$-isobar
in the nucleus.
Adler's
PCAC theorem is used to connect the cross section of the $(\nu_l,l)$
reaction to that of the pion induced reaction.
The
measured cross section of the later is used (as input) to avoid the
uncertainties in its calculated cross section.
The
calculated results show that the distinct shift and broadening of the
$\Delta$-peak in the nucleus relative to proton in the inclusive
$(\nu_l,l)$ reaction, and those in the nucleus increases with the size
of nucleus.
It
appears because of the exclusive and semi-exclusive channels which occur in
the nucleus. These channels do not exist for proton.
The
modification of the $\Delta$-peak in the neutrino nucleus reaction, studied
using Adler's PCAC  theorem, is resemble to that seen in the hadron (pion)
nucleus reaction which occurs at the periphery of the nucleus. It is unlike
to that occurs in the gamma or electron induced nuclear reaction, though
the nuclear reaction based on the electro-weak interaction takes place
throughout the volume of the nucleus.

\section{Acknowledgement}

The author takes the opportunity to thank an anonymous referee for giving
valuable comments which helped to improve the quality of the paper.



\begin{figure}[h]
\begin{center}
\centerline {\vbox {
\psfig{figure=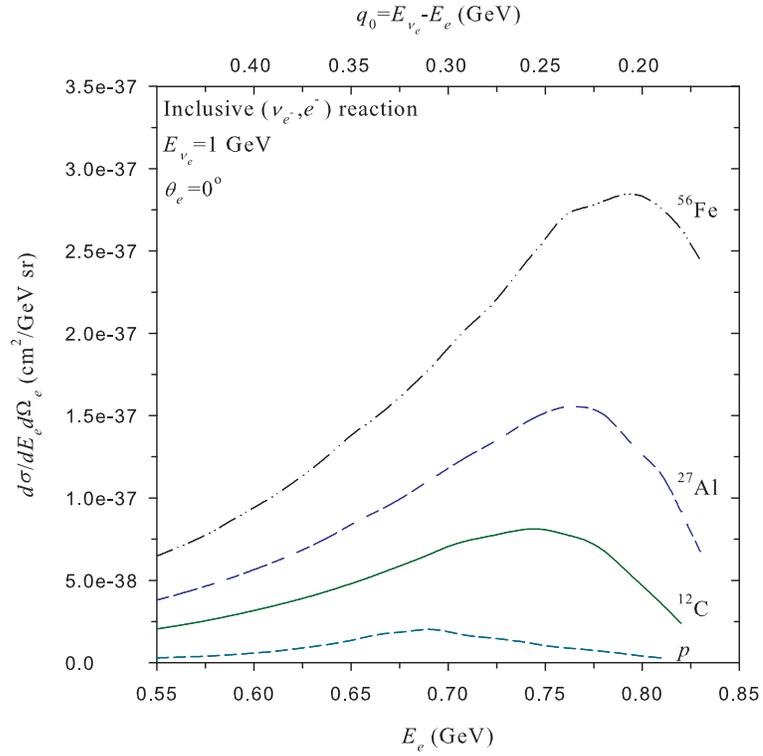,height=10.0 cm,width=10.0 cm}
}}
\caption{
(color online).
The shift and broadening of the $\Delta$-peak in the inclusive
$(\nu_{e^-},e^-)$ reaction on nuclei relative to proton. The modification
of the peak increases with the size of the nucleus.
}              
\label{fg1}
\end{center}
\end{figure}

\begin{figure}[h]
\begin{center}
\centerline {\vbox {
\psfig{figure=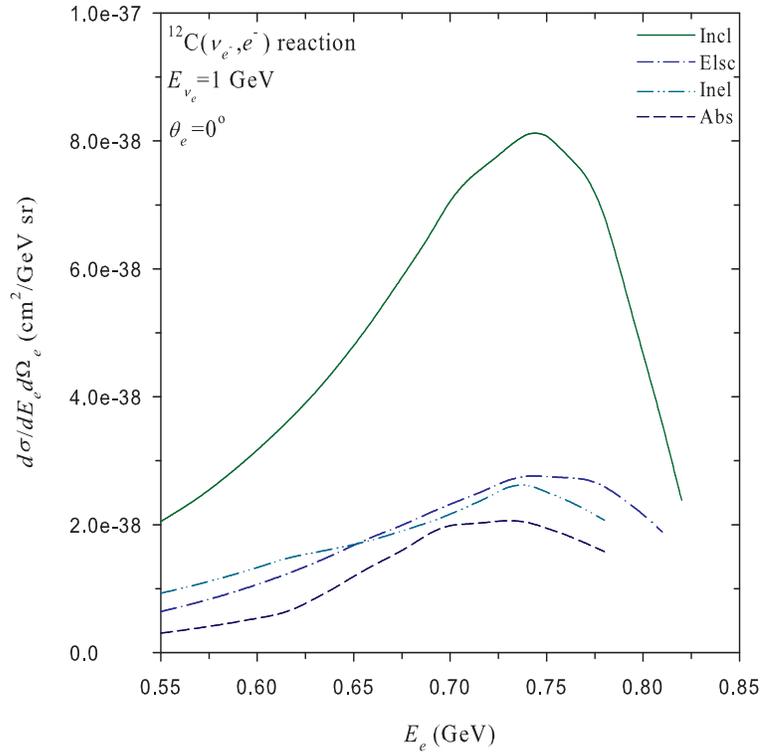,height=10.0 cm,width=10.0 cm}
}}
\caption{
(color online).
The cross sections of the inclusive and exclusive $(\nu_{e^-},e^-)$
reactions on $^{12}$C nucleus. The labels appearing in the figure represent
various reactions: ``Incl'' (inclusive reaction), ``Elsc'' (elastic pion
production), ``Inel'' (inelastic pion production) and ``Abs'' (no pion in
the final state).
}              
\label{fg2}
\end{center}
\end{figure}

\begin{figure}[h]
\begin{center}
\centerline {\vbox {
\psfig{figure=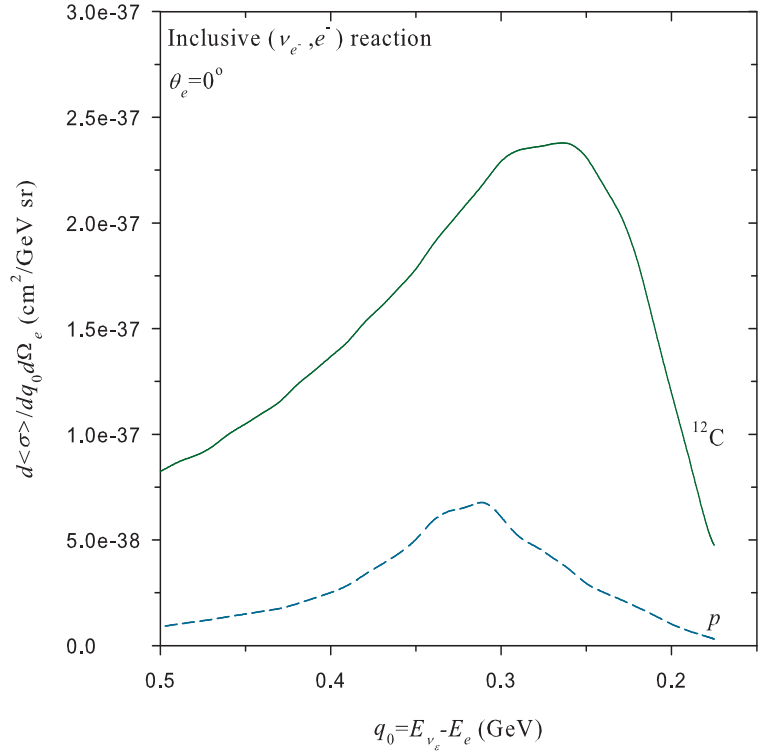,height=10.0 cm,width=10.0 cm}
}}
\caption{
(color online).
The flux averaged differential cross section for the forward going
ejectile in the inclusive $(\nu_{e^-},e^-)$ reaction in the
$\Delta$-excitation region. $q_0=E_{\nu_e}-E_e$ is the energy transfer to
the nucleus. The modification of the $\Delta$-peak is distinctly visible
in the figure.
}              
\label{fg3}
\end{center}
\end{figure}

\begin{figure}[h]
\begin{center}
\centerline {\vbox {
\psfig{figure=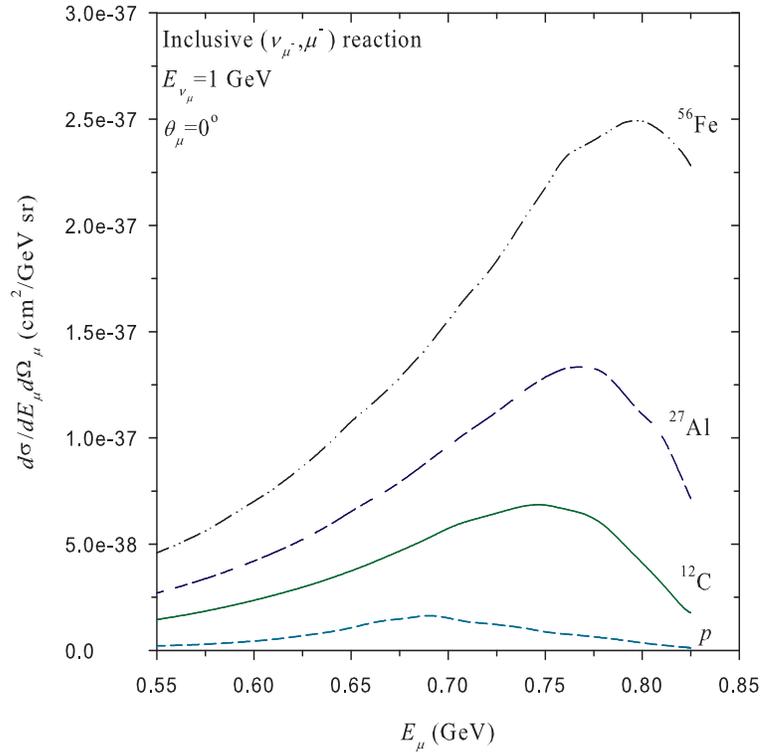,height=10.0 cm,width=10.0 cm}
}}
\caption{
(color online).
Same as those presented in Fig.~\ref{fg1} but for the $(\nu_{\mu^-},\mu^-)$
reaction. The modification of $\Delta$-peak is qualitatively similar to
that in Fig.~\ref{fg1}.
}              
\label{fg4}
\end{center}
\end{figure}

\begin{figure}[h]
\begin{center}
\centerline {\vbox {
\psfig{figure=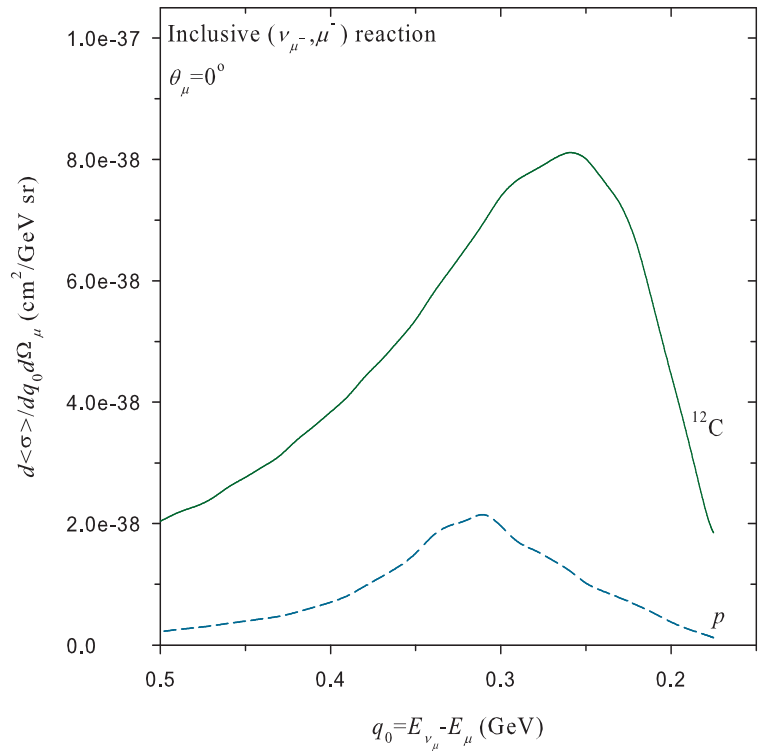,height=10.0 cm,width=10.0 cm}
}}
\caption{
(color online).
Same as those presented in Fig.~\ref{fg3} but for the $(\nu_{\mu^-},\mu^-)$
reaction (see text).
}              
\label{fg5}
\end{center}
\end{figure}

{\bf Figure Captions}
\begin{enumerate}

\item
(color online).
The shift and broadening of the $\Delta$-peak in the inclusive
$(\nu_{e^-},e^-)$ reaction on nuclei relative to proton. The modification
of the peak increases with the size of the nucleus.

\item
(color online).
The cross sections of the inclusive and exclusive $(\nu_{e^-},e^-)$
reactions on $^{12}$C nucleus. The labels appearing in the figure represent
various reactions: ``Incl'' (inclusive reaction), ``Elsc'' (elastic pion
production), ``Inel'' (inelastic pion production) and ``Abs'' (no pion in
the final state).

\item
(color online).
The flux averaged differential cross section for the forward going
ejectile in the inclusive $(\nu_{e^-},e^-)$ reaction in the
$\Delta$-excitation region. $q_0=E_{\nu_e}-E_e$ is the energy transfer to
the nucleus. The modification of the $\Delta$-peak is distinctly visible
in the figure.

\item
(color online).
Same as those presented in Fig.~\ref{fg1} but for the $(\nu_{\mu^-},\mu^-)$
reaction. The modification of $\Delta$-peak is qualitatively similar to
that in Fig.~\ref{fg1}.

\item
(color online).
Same as those presented in Fig.~\ref{fg3} but for the $(\nu_{\mu^-},\mu^-)$
reaction (see text).

\end{enumerate}

\newpage

\newpage

\newpage

\newpage

\newpage

\end{document}